\documentclass[english,aps,preprint]{revtex4-1}
\usepackage[T1]{fontenc}
\usepackage[latin9]{inputenc}
\usepackage[letterpaper]{geometry}
\usepackage{amstext}
\usepackage{amssymb}
\usepackage{graphicx}
\usepackage{esint}

\makeatother

\usepackage{babel}
\begin{document}

\title{Propagator with Positive Cosmological Constant in the 3D Euclidian
Quantum Gravity Toy Model}

\author{William E. Bunting}

\affiliation{California Institute of Technology, Pasadena, CA 91125, USA}

\author{Carlo Rovelli}

\affiliation{Aix Marseille Université, CNRS, CPT, UMR 7332, 13288 Marseille, France.}
\begin{abstract}
We study the propagator on a single tetrahedron in a three dimensional
toy model of quantum gravity with positive cosmological constant.
The cosmological constant is included in the model via q-deformation
of the spatial symmetry algebra, that is, we use the Tuarev-Viro amplitude.
The expected repulsive effect of dark energy is recovered in numerical
and analytic calculations of the propagator at large scales comparable
to the infrared cutoff. However, due to the simplicity of the model
we do not obtain the exact Newton limit of the propagator. This is
a first step toward the similar calculation in the full 3+1 dimensional
theory with larger numbers of simplicies. 
\end{abstract}
\maketitle

\section{Introduction\label{sec:Introduction}}

This paper extends work \cite{Speziale_2006a,Speziale_2006b,Bonzom2008507}
evaluating quantum gravity two-point functions on a single tetrahedron
in three euclidean dimensions to the case with positive cosmological
constant. At long distance scales the two point function for a quantum
field theory gives the Newton force associated to that theory's gauge
boson. In our toy model, since we only have one tetrahedron and we
peak our state around an equilateral configuration of that tetrahedron
we do not expect to reproduce the exact Newton limit of 3D gravity
with a cosmological constant. However, we still do find an asymptotically
repulsive force associated to the dark energy. 

We take the inclusion of the cosmological constant to correspond to
a deformation of the $SU(2)$ spatial rotation symmetry. This quantum
gravity toy model is known as the q-deformed Ponzano-Regge model or
the Tuarev-Viro model. We expect two key differences in the propagator
from the Ponzano-Regge model. First, in the case where the tetrahedra
is large compared with the infrared cutoff imposed by the cosmological
constant, the sums will be cut off by the deformation and not the
triangular inequalities. This feature of the quantum algebra is essential
in four dimensions where we cannot escape so-called ``bubble'' configurations
of 2-complexes, without an infrared cutoff these transition amplitudes
would diverge. Second, in the case where the size of the tetrahedron
is smaller than the cutoff, the modification will simply affect the
asymptotics of the two point function via the addition of a volume
term to the Regge action.

In this paper we first compute the Tuarev-Viro propagator numerically
in Section \ref{sec:Method-and-Results}. Then we compare the numerical
computations with an analytic calculation of the propagator asymptotics
in Section \ref{sec:Discussion-and-Analysis}. These two calculations
are found to match strongly in an easily computable regime, that is
for an unrealistically large cosmological constant. However, the agreement
is expected to persist as the cosmological constant is taken smaller.

\section{Method and Results\label{sec:Method-and-Results}}

\begin{figure}
\noindent \begin{centering}
\includegraphics[scale=0.65]{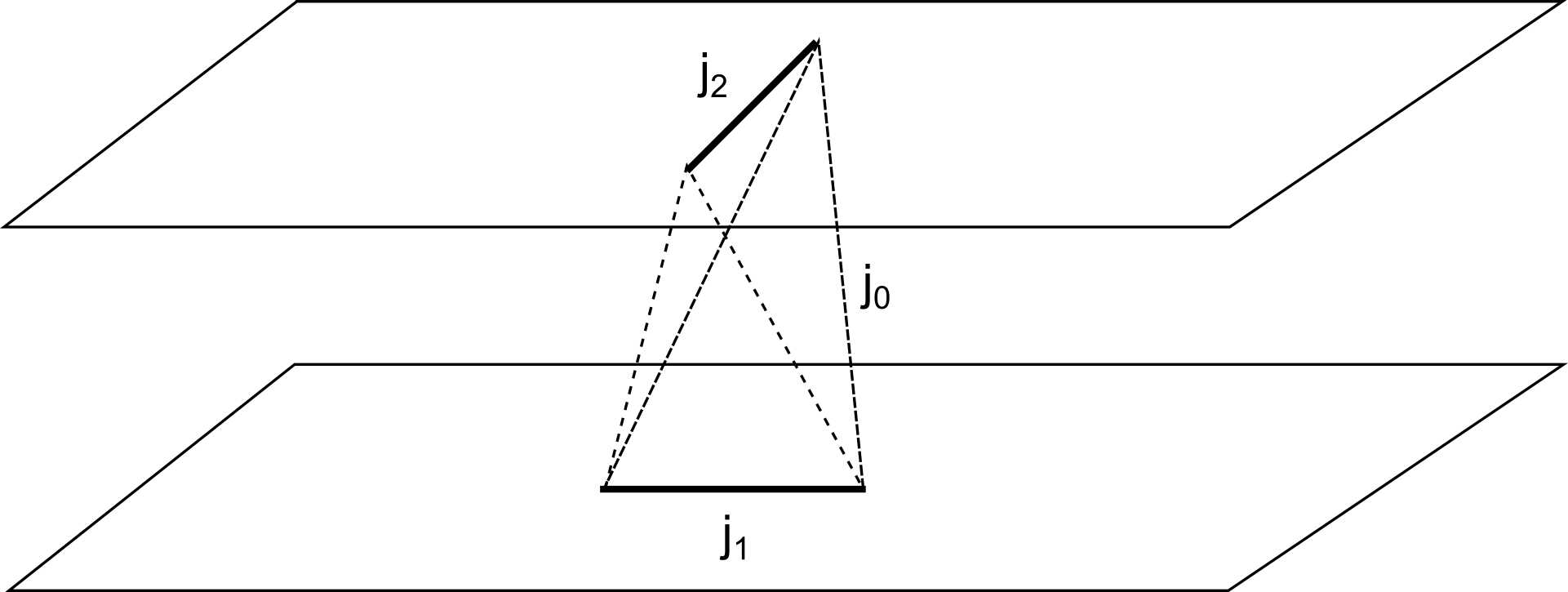}
\par\end{centering}

\caption{Depiction of the tetrahedron state used in this paper. The the four
edges of the tetrahedron that go between the boundary slices are fixed
to a value of $j_{0}$ and can be thought to correspond to the time.
We then study the two point function between the length operators
acting on each of the two edges on the slices.\label{fig:speziale_fig}}
\end{figure}

This paper studies the correlator between two edges on a single tetrahedron.
We fix four edges around the tetrahedron to $j_{0}$ which can be
thought of as a time, thus $j_{1}$ and $j_{2}$ are lengths on two
different time slices of the spacetime, see Figure \ref{fig:speziale_fig}.
We will study the modulus of the propagator $|\mathcal{P}|$ as a
function of the distance $j_{0}$ and the infrared cutoff $j_{\text{max}}$. 

\begin{figure}
\noindent \begin{centering}
\includegraphics[scale=0.6]{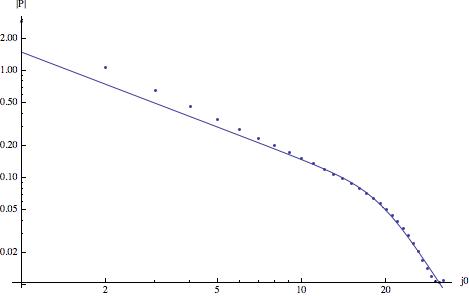}
\par\end{centering}

\caption{Plot of the q-deformed propagator calculated for $j_{\text{max}}=65$
and various values of $j_{0}$. The asymptotic behavior is clearly
modified by the presence of the cosmological constant via a repulsion
at large distances. This repulsion is well captured in the analytic
calculation of the curve in the figure. Additionally, in the intermediate
regime where $1\ll j_{0}\ll j_{\text{max}}/2$ we have agreement with
the Speziale asymptotics ie. $|\mathcal{P}|\sim3/2j_{0}$. \label{fig:biglambda}}
\end{figure}

Because in the 3D case all of the calculations can be done in a gauge
where they all give zero, we clearly have an issue of gauge choice.
We follow Speziale in by picking a Coulomb-like gauge where the field
operators have non-trivial projections along a bone (edge of a triangle
in the triangulation) \cite{Speziale_2006a}. This choice will produce
a calculation similar to the one that can be done in the 4D spinfoam
model \cite{Bianchi_Rovelli_Speziale_2006}. We choose our operator
notations different than Speziale, but in line with more recent work
in quantum gravity correlation functions \cite{Alesci_Rovelli_2013,Rovelli_Zhang_3pt_2011}.
The two point function is:
\begin{equation}
\mathcal{P}_{nm}^{abcd}:=\frac{\langle W|\mathcal{P}_{n}^{ab}\mathcal{P}_{m}^{cd}|\psi_{\Sigma}\rangle_{q}}{\langle W|\psi_{\Sigma}\rangle_{q}},\label{eq:2pointEquation}
\end{equation}
where instead of the usual metric field insertions we have the operators
perturbed around flat space.
\begin{eqnarray*}
\mathcal{P}_{n}^{ab}|s\rangle & := & l_{a,n}^{\mu}l_{b,n}^{\nu}h_{\mu\nu}(x)|s\rangle\\
 & = & l_{a,n}^{\mu}l_{b,n}^{\nu}(g_{\mu\nu}(x)-\delta_{\mu\nu})|s\rangle\\
 & = & \left(l_{p}^{2}G_{n}^{ab}-l_{p}^{2}C_{q}^{2}(j_{0})\right)|s\rangle
\end{eqnarray*}
Above $G_{n}^{ab}:=\vec{L}_{n}^{a}\cdot\vec{L}_{n}^{b}$ are the Penrose
operators acting on links that go between nodes $a$ and $n$, and
$b$ and $n$ in the spin network state $|s\rangle$, $C_{q}$ is
the $SU_{q}(2)$ casimir defined as $C_{q}(j):=\sqrt{[j][j+1]+1/4}$,
$W$ is the Tuarev-Viro transition amplitude and $\psi_{\Sigma}$
is a boundary state peaked on an equilateral tetrahedron. For a tetrahedron
the spin network state is also a tetrahedron dual to the original
one where there are nodes on all the faces. We would like to choose
the node labels such that we are calculating the correlator between
the two edges labeled by $j_{1}$ and $j_{2}$. For instance, we could
label the four nodes such that face $1$ opposes face $3$ and face
$2$ opposes face $4$. Then we would be interested in $\mathcal{P}_{34}^{1122}$.
We calculate this by evaluating equation \ref{eq:2pointEquation}
with the q-deformed state and transition amplitude, ie.:
\begin{equation}
\mathcal{P}_{34}^{1122}=\frac{1}{j_{0}^{4}l_{p}^{4}\mathcal{N}}\sum_{j_{1},j_{2}}^{2j_{0}}W_{q}(j_{1},j_{2};j_{0})\psi_{\Sigma}^{q}(j_{1},j_{2};j_{\text{max}})l_{p}^{4}\left(C_{q}^{2}(j_{1})-C_{q}^{2}(j_{0})\right)\left(C_{q}^{2}(j_{2})-C_{q}^{2}(j_{0})\right),\label{eq:propagator}
\end{equation}
where we have from the Tuarev-Viro model that the transition amplitude
for a single tetrahedron is just a quantum 6j-symbol:
\[
W_{q}(j_{1},j_{2};j_{0})=\left\{ \begin{array}{ccc}
j_{1} & j_{0} & j_{0}\\
j_{2} & j_{0} & j_{0}
\end{array}\right\} _{q},
\]
we take the deformation parameter to be a root of unity $q=\exp(i\sqrt{\Lambda} \hbar G)$
where $\Lambda$ is the cosmological constant that fixes the infrared
cutoff $j_{\text{max}}$ as $\sqrt{\Lambda}=\pi/(2j_{\text{max}}+1)$
\cite{Tada_Mizoguchi1992}. The boundary state is taken to be:
\[
\psi_{\Sigma}^{q}=\frac{1}{N}\exp\left(-\frac{\alpha}{2}\sum_{i}(j_{i}-j_{0})^{2}+i\theta\sum_{i}\left(j_{i}+\frac{1}{2}\right)-i\frac{\sqrt{\Lambda}j_{0}^{2}}{12\sqrt{2}}\sum_{i}(j_{i}-j_{0})\right).
\]
The factor with $\Lambda$ in this state is included so that the asymptotics
do not have oscillatory terms in the limit as $j_{0}\rightarrow\infty$
that come from the addition of the cosmological constant term to the
Regge action. Numerical evaluation of equation \ref{eq:propagator}
gives the main result of the paper, see figure \ref{fig:biglambda}.
Here we can see that at small scales $j_{0}\approx1$ we have the
same quantum deviations as Speziale from the Newtonian limit. In the
range where we are not too close to the cutoff and not too small lengths
ie. $1\ll j_{0}\ll j_{\text{max}}/2$ we have behavior similar to
the Newtonian limit without cosmological constant $|\mathcal{P}|\approx3/2j_{0}$.
Lastly, when the spins get close to the infrared cutoff we have a
repulsion representative of the repulsive force of dark energy. 

\section{Discussion and Analysis\label{sec:Discussion-and-Analysis}}

We would like to analyze analytically the asymptotics of this toy
model to see if we can match the numerical result. The asymptotics
of $\{6j\}_{q}$ symbols are given by a cosine of the Regge action
with a volume term \cite{Woodward_2003,Roberts_2002}. 

\[
\{6j\}_{q}\sim\frac{2\pi\cos\left(S_{R}[j_{e}]-\sqrt{\Lambda}\text{Vol}(\tau_{S})+\frac{\pi}{4}\right)}{r^{3/2}G^{1/4}}
\]
The volume term is the volume of a spherical tetrahedron. We will
approximate this as a flat tetrahedron and expand the volume in $\delta j_{1}$
and $\delta j_{2}$ as:

\[
\text{Vol}(\tau_{S})\approx\text{Vol}(\tau_{E})\approx\frac{(1+2j_{0})^{3}}{48\sqrt{2}}+\frac{(1+2j_{0})^{2}}{48\sqrt{2}}\left(\delta j_{1}+\delta j_{2}\right)-\frac{7(1+2j_{0})}{96\sqrt{2}}\left(\delta j_{1}^{2}+\delta j_{2}^{2}\right)-\frac{(1+2j_{0})\delta j_{1}\delta j_{2}}{48\sqrt{2}}.
\]
The factor of $2\pi/r^{3/2}G^{1/4}$ is approximately independent
of $j_{1}$ and $j_{2}$ therefore it will cancel with the same factor
in the normalization. Also in the large spin limit the difference
of the Casimir operators will go like $C^{2}(j_{1})-C^{2}(j_{0})\sim2j_{0}\delta j_{1}$.
We then arrive at an asymptotic formula for the propagator:

\[
|\mathcal{P}|\sim\frac{4}{j_{0}^{2}\mathcal{N}}\sum_{j_{1},j_{2}}^{2j_{0}}\delta j_{1}\delta j_{2}\exp\left[-iS_{R}[j_{e}]+\sqrt{\Lambda}i\text{Vol}(\tau_{E})-i\frac{\pi}{4}-\frac{\alpha}{2}\sum_{i}\delta j_{i}^{2}+i\theta\sum_{i}\left(j_{i}+\frac{1}{2}\right)-\frac{i\sqrt{\Lambda}j_{0}^{2}}{12\sqrt{2}}\sum_{i}\delta j_{i}\right],
\]
where we have omitted the other exponential term that is rapidly oscillating
in $j_{1}$ and $j_{2}$. Expanding action and the volume terms around
the equilateral configuration, and canceling factors independent of
$j_{1}$ and $j_{2}$ with the same ones in the normalization we obtain:
\[
|\mathcal{P}|\sim\frac{4}{j_{0}^{2}\mathcal{N}}\sum_{j_{1},j_{2}}^{2j_{0}}\delta j_{1}\delta j_{2}\exp\left[-\frac{i\sqrt{\Lambda}}{2}\left(\frac{7j_{0}}{24\sqrt{2}}\left(\delta j_{1}^{2}+\delta j_{2}^{2}\right)+\frac{j_{0}\delta j_{1}\delta j_{2}}{12\sqrt{2}}\right)-\frac{i}{2}\sum_{i,k}G_{ik}\delta j_{i}\delta j_{k}-\frac{\alpha}{2}\sum_{i=1}^{2}\delta j_{i}^{2}\right],
\]
where $G_{ik}=\partial\theta_{i}/\partial j_{k}|_{j_{e}=j_{0}}$ is
the matrix of derivatives of the dihedral angles with respect to the
edge lengths evaluated for an equilateral tetrahedron. Passing to
continuous variables $z=\delta j_{1}$ and $dz=dj_{1}$ we obtain
the gaussian integral:
\[
\mathcal{P}\sim\frac{4}{j_{0}^{2}\mathcal{N}}\int d^{2}z\; z_{1}z_{2}\exp\left[-\frac{1}{2}z_{i}A_{ik}z_{k}\right]=\frac{4}{j_{0}^{2}}\left(A^{-1}\right)_{12},
\]
where $A$ is the matrix of coefficients of $\delta j_{i}\delta j_{k}$:
\[
A_{ik}=\frac{4}{3j_{0}}\left(\begin{array}{cc}
1+i\frac{\sqrt{2}}{4} & i\frac{3\sqrt{2}}{4}\\
i\frac{3\sqrt{2}}{4} & 1+i\frac{\sqrt{2}}{4}
\end{array}\right)+\frac{i\sqrt{\Lambda}(j_{0}+1/2)}{24\sqrt{2}}\left(\begin{array}{cc}
7 & 1\\
1 & 7
\end{array}\right),
\]
where we have set $\alpha=4/3j_{0}$ which is compatible with the
requirement that the state be peaked on the intrinsic and extrinsic
geometry \cite{Speziale_2006a}. From this, we have the full expression
for the asymptotics of the propagator modulus:

\[
|\mathcal{P}|\sim\frac{4\sqrt{6}(96+J_{\Lambda})^{2}}{j_{0}}\frac{1}{\sqrt{2^{17}\cdot3+J_{\Lambda}(2^{13}+J_{\Lambda}(128+J_{\Lambda}(32+3J_{\Lambda})))}}
\]
where $J_{\Lambda}:=j_{0}(1+2j_{0})\sqrt{\Lambda}$. The plot of this
analytic expression versus the numerical calculation is shown in Figure
\ref{fig:biglambda} and the agreement is seen to be very good. If
we expand this expression about $\Lambda=0$ it becomes more clear
that a repulsion is the dominating correction to the Speziale result. 

\[
|\mathcal{P}|\sim\frac{3}{2j_{0}}-\frac{1}{1536}(j_{0}+1/2)^{2}j_{0}\Lambda+O(\Lambda^{3/2})
\]
We obtain an expression with a term proportional to the cosmological
constant times the volume of the tetrahedron. While this differs from
what we might expect from the Newton law, we should remember that
this is a simplified model where there is only one simplex that spans
the cosmological distance, and it is peaked on an equilateral configuration.
Further investigations could attempt a similar calculation where there
was a small length scale introduced to make an elongated tetrahedron
instead of an equilateral one, which would better capture the true
propagator physics. Despite these caveats we still observe the characteristic
repulsion of a positive cosmological constant.

\section{Summary and Conclusions}

We have shown that the inclusion of a cosmological constant in a 3D
euclidean toy model of quantum gravity on a single tetrahedron reproduces
the expected qualitative behavior near the infrared cutoff. Namely,
we expect to have an additional repulsive force on large distance
scales. The effect is reproduced in both numerical calculations and
an analytic evaluation of the propagator asymptotics. This work also
strengthens the argument for the interpretation of the cosmological
constant as a deformation parameter in the theory. However, as mentioned
in the introduction, we do not produce the exact form of the Newton
propagator, which is thought to be due to the simplicity of the model. 

This result can be taken as a first step toward performing a similar
calculation in 3+1 dimensions. Given that there are already asymptotic
calculations of the propagator \cite{Bianchi_Rovelli_Speziale_2006}
and the $SU(2)$ symmetry is still the relevant one this may not be
too difficult for one 4-simplex. Another direction this work could
be extended is to multiple tetrahedra. The numerical computations
become significantly more difficult as the number of simplicies increase,
however one could probably check the analytic results to see if the
repulsion persists.

\section{Acknowledgements}

The authors would like to thank Simone Speziale and François Collet
for helpful discussions. WB is funded by the Caltech Summer Undergraduate
Research Fellowship Program. 

\bibliographystyle{apsrev}
\bibliography{3D_QG_qDef_Paper}

\end{document}